\documentclass[11pt]{article}
\pdfoutput=1
 \usepackage{mciteplus}
 \usepackage{tikz}
 \usepackage{color}
 \definecolor{darkblue}{rgb}{0.1,0.1,.7}
 \usepackage[colorlinks, linkcolor=darkblue, citecolor=darkblue, urlcolor=darkblue, linktocpage,hyperfootnotes=false]{hyperref} 
\usepackage{epsfig}
\usepackage{graphicx}
\usepackage{cite}
\usepackage{amsfonts}
\usepackage{amssymb}
\usepackage{bm}
\usepackage{latexsym}
\usepackage{amsmath}
\numberwithin{equation}{section}
\def\bq{\begin{quote}}
\def\eq{\end{quote}}
 at 10truept

\newcommand{\calo}{{\cal O}}
\newcommand{\beq}{\begin{equation}}
\newcommand{\eeq}{\end{equation}}
\newcommand{\beqa}{\begin{eqnarray}}
\newcommand{\eeqa}{\end{eqnarray}}
\newcommand{\bea}{\begin{eqnarray}}
\newcommand{\eea}{\end{eqnarray}}

\def\roughly#1{\raise.3ex\hbox{$#1$\kern-.75em\lower1ex\hbox{$\sim$}}}
\newcommand{\Tr}{{\rm Tr}}

\begin{document}

\thispagestyle{empty}
\begin{titlepage}
  \bigskip

  \bigskip\bigskip

  \bigskip

\begin{center}
{\Large \bf {Wormhole calculus, replicas, and entropies }}
    \bigskip
\bigskip
\end{center}

  \begin{center}

 \rm {Steven B. Giddings\footnote{\texttt{giddings@ucsb.edu}} and Gustavo J. Turiaci\footnote{\texttt{turiaci@ucsb.edu}}}
  \bigskip \rm
\bigskip

{Department of Physics, University of California, Santa Barbara, CA 93106, USA}  \\
\rm

  \bigskip \rm
\bigskip
 
\rm

\bigskip
\bigskip

  \end{center}

\vspace{3cm}
  \begin{abstract}

We investigate contributions of spacetime wormholes, describing baby universe emission and absorption, to calculations of entropies and correlation functions, for example those based on the replica method.  We find that the rules of the ``wormhole calculus," developed in the 1980s, together with standard quantum mechanical prescriptions for computing entropies and correlators, imply definite rules for {\it limited} patterns of connection between replica factors in simple calculations.  These results stand in contrast with assumptions that all topologies connecting replicas should be summed over, and call into question the explanation for the latter. In a ``free" approximation baby universes introduce probability distributions for coupling constants, and we review and extend arguments that successive experiments in a ``parent" universe increasingly precisely fix such couplings, resulting in ultimately pure evolution.  Once this has happened, the nontrivial question remains of how topology-changing effects can modify the standard description of black hole information loss.

 \medskip
  \noindent
  \end{abstract}
\bigskip \bigskip \bigskip 

  \end{titlepage}

\section{Introduction}

Nontrivial spacetime topologies, and in particular change in the topology of space, have long been considered to be a possible feature of dynamical gravity.  Topology-changing processes were particularly intensively studied in the late 1980s, in the context of  the question of their contribution to possible loss of quantum coherence\cite{Lavrelashvili:1987jg,Hawking:1987mz,GiStAx,Cole,GiStInc}.  Specifically, one can consider processes where space branches into two disconnected components; one of these may typically be comparatively small, and was called a ``baby universe" (BU).  In the ``free BU" approximation where multiple BUs can be emitted, or rejoin, a bigger ``parent universe,"  but where the BUs don't interact or create other large universes, it was found that the leading effect of such processes is {\it not} to induce an ongoing loss of quantum coherence\cite{Cole,GiStInc}.\footnote{Effects beyond this approximation were discussed in \cite{GiSt3Q}, and recently in \cite{Marolf:2020xie}.}  Instead, these processes lead to an effective probability distribution for coupling constants that multiply operators describing the effect of the BUs on the fields in the parent universe.  

There has been a recent resurgence of interest in topology change, arising from suggestions that nontrivial topologies may help explain how black hole evolution can be reconciled with unitary quantum mechanical evolution \cite{Saad:2019lba, Penington:2019kki, Almheiri:2019qdq}.\footnote{For 
earlier work in this direction, see \cite{Polchinski:1994zs,AstroLH}.  For a different but possibly related approach see \cite{NVU,BHQU}.}  Specifically, \cite{Penington:2019kki, Almheiri:2019qdq} have argued that nontrivial topological contributions can produce expressions for BH entropies that behave as expected for unitary evolution\cite{Pageone,Page:1993wv}.  This work builds on earlier discussion\cite{Penington:2019npb, Almheiri:2019psf, Almheiri:2019hni} about the role of quantum extremal surfaces, and that of \cite{Saad:2019lba} on topologies and ensembles of couplings in Jackiw-Teitelboim gravity (see also the related work \cite{Maldacena:2019cbz, Cotler:2019nbi, Stanford:2019vob, Saad:2019pqd, Blommaert:2019wfy}).  The topologies studied in \cite{Almheiri:2019qdq, Penington:2019kki} involve spacetime wormhole connections, but of a somewhat different kind than those studied in the 1980s.  Specifically, entropies are calculated by the replica method\cite{CaWi}, in which multiple copies of the spacetime geometry are considered.  One then makes the Ansatz that wormholes, or more general nontrivial topologies, connect these replicas.  While the replica wormhole contributions have not yet been shown to correspond to quantum amplitudes describing unitary evolution, they have been argued to produce entropy formulas that reflect unitary behavior, giving an appropriate form of a ``Page curve"\cite{Pageone,Page:1993wv}.

The obvious possible connection between replica wormholes and the spacetime wormholes considered previously was noted in \cite{Penington:2019kki, Almheiri:2019qdq}, and further developed in \cite{Marolf:2020xie}.  However,  an important question in the discussion is to better understand the precise connection, and to test and understand the correct rules for replica calculations in the presence of euclidean wormholes/BU emission.  Specifically,  \cite{GiStInc},\cite{Cole} previously developed a set of rules for incorporating topology change, respecting certain general quantum properties such as the composition of amplitudes; this is sometimes called the ``wormhole calculus."  Given the wormhole calculus, one can then perform standard quantum-mechanical calculations -- such as of entropies, {\it e.g.} using replicas -- and ask what the combined set of rules tells us about the contribution of nontrivial topologies connecting replicas, and  regarding the question of summing over all such replica geometries.

That is one of the goals of this paper.  Specifically, we find that the previously-developed  rules of the wormhole calculus, which have been well studied in a framework consistent with quantum mechanics,  together with basic quantum-mechanical rules for computing entropies, imply {\it specific limited patterns of wormhole connections in replica geometries.}  These do not include sums over all connections between replicas. This runs contrary to the prevalent Ansatz that one should generally sum over all such replica topologies\cite{Penington:2019kki, Almheiri:2019qdq}, and calls into question the meaning of calculations based on such a sum.   Specifically this suggests that if there is a role for replica wormholes in certain calculations, it needs to be more carefully understood; alternatively it may also be that including such contributions represents a modification of usual quantum-mechanical rules for calculating entropies, or somehow gives an effective description summarizing the contribution of other effects.

In outline, the next section gives a brief review of the wormhole calculus.  Section three then turns to the question of calculating some simple entropies, as well as correlators, in the presence of nontrivial spacetime topologies and ensembles of BUs, showing that the wormhole calculus together with the usual rules dictate only certain patterns of wormhole connections between replicas.  Section four discusses a related question, namely that of understanding the effects of BUs as providing a probability distribution for coupling constants, and the way in which subsequent experiments determine these couplings; this provides a generalization of the analysis of \cite{Cole,GiStInc} of these questions.  Section five closes with some further discussion.

\section{Review of the wormhole calculus}

We begin by reviewing the basics of the wormhole calculus, developed in \cite{GiStInc},\cite{Cole}. This was based on assuming the existence of topology-changing interactions in which a universe can split, emitting a disconnected baby universe (BU).  A simple instanton describing such processes, in the presence of a massless axionic field, was found in \cite{GiStAx}; similar processes were also considered by \cite{Hawking:1987mz, Lavrelashvili:1987jg, Klebanov:1988eh, Rubakov:1988jf, Rey:1989mg, ColeCC,Polchinski:1994zs, Maldacena:2004rf, ArkaniHamed:2007js}.  

Specifically, suppose that we work in the free BU approximation where BUs can be emitted and absorbed by a single parent universe, but do not interact among themselves or create other large universes; going beyond this approximation can be described in a third-quantized framework\cite{GiSt3Q}.  For simplicity, consider the case where the parent universe has an asymptotic region where time can be defined, such as asymptotically flat or AdS space.  Then, we can consider finite-time transitions between states of the parent universe, but at the same time there can be transitions in the number of BUs.

\begin{figure}[h!]
\begin{center}
  \begin{tikzpicture}[scale=1]
\pgftext{\includegraphics[scale=0.5]{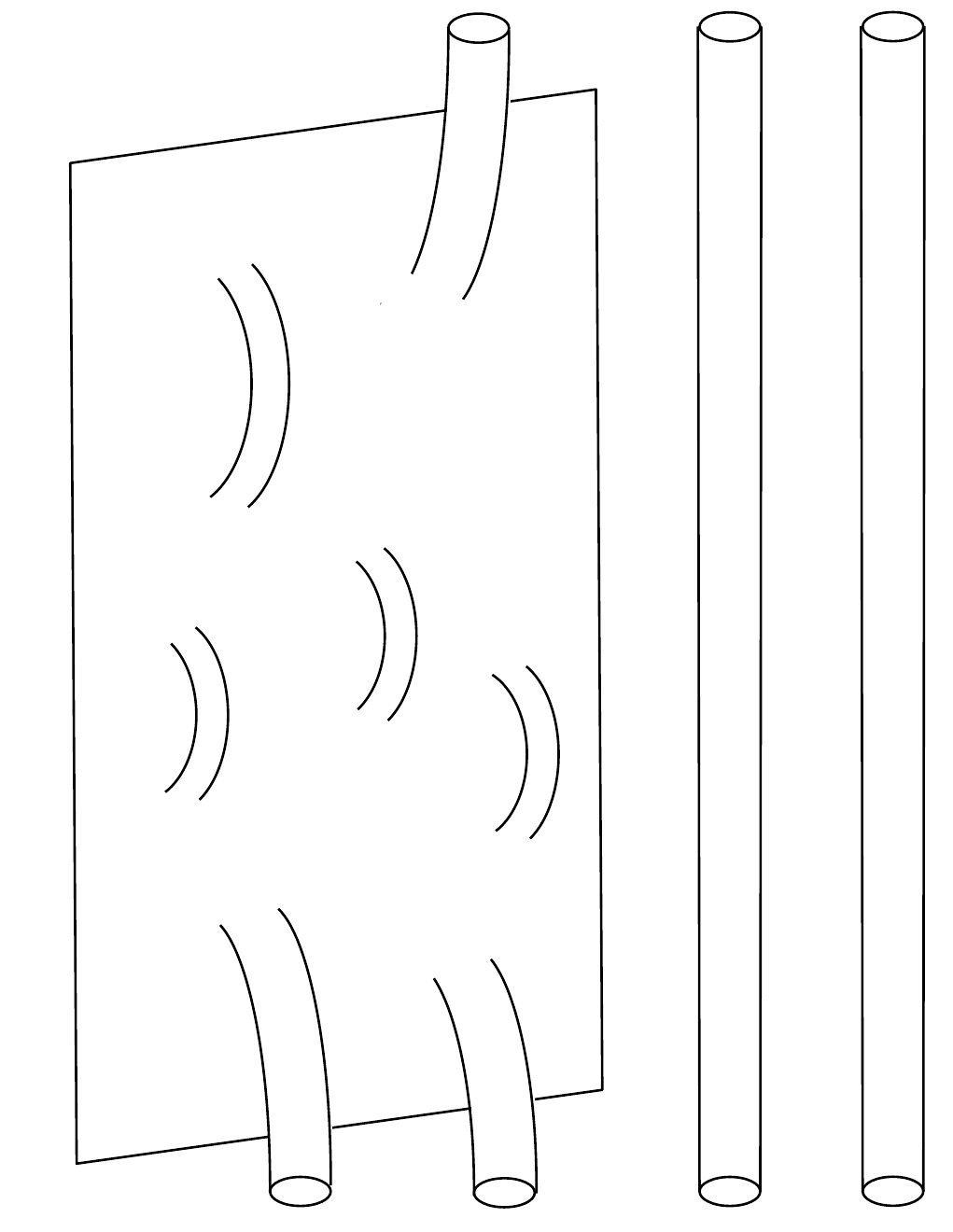}} at (0,0);
\draw (3.9,-3) node  {\small $|\psi_i, n_i = 4\rangle$};
\draw (3.9,2.9) node  {\small $|\psi_f, n_f = 3\rangle$};
  \end{tikzpicture}
 \caption{\label{fig:whamp} To compute the amplitude to go from an initial state of the parent universe, plus some number of BUs, to a similar final state, we integrate over all intermediate geometries.  In this figure we sketch one particular geometry contributing to the transition amplitude between  four and three BUs.}
\end{center}
\end{figure}

In general, the BUs can have different internal states, but for simplicity consider the case where there is a single internal state, or 
 ``species," of BU.  Then, one can consider transitions between an initial state of the parent universe, together with some number of BUs, and a final state of the parent universe together with some typically different number of BUs.  The amplitudes for such processes can be calculated by summing over geometries such as in fig.~\ref{fig:whamp}, in analogy to other instanton sums in physics.

As shown in \cite{Cole,GiStInc}, at scales large as compared to the typical BU size (which may be set by a microscopic scale), these amplitudes can be reproduced from a simple hamiltonian.  This takes the form
\beq\label{BUham}
H= H(\phi_i) + \int d^3x\, \calo(x) (a+a^\dagger)\ .
\eeq
Here $\phi_i$ are the fields on the parent universe (which may also include the metric), $H(\phi_i)$ is their hamiltonian, and $\calo(x)$ is an operator that describes the effect of the BU emission on these fields.  The operators $a^\dagger$ and $a$ act on a BU Fock space, to create/annihilate BUs; for example, an $n$ BU state
is given by
\beq
|n\rangle = \frac{1}{\sqrt{n!}} (a^\dagger)^n |0\rangle
\eeq
where $|0\rangle$ is the BU vacuum.  The form of the hamiltonian \eqref{BUham} is dictated by various considerations: the fact that BU emission conserves energy/momentum, since BUs are closed and carry no net energy/momentum, indistinguishability of BUs, and the requirement that the basic amplitudes, of the form
\beq
\langle \psi_f,n_f|e^{-i HT}|\psi_i,n_i\rangle\ ,
\eeq
satisfy a composition law,
\beq
\sum_{n, \psi_I}\langle\psi_f,n_f |e^{-iHT_2}|\psi_I,n\rangle\langle \psi_I,n|e^{-iHT_1}|\psi_i, n_i\rangle = \langle\psi_f,n_f |e^{-iH(T_1+T_2)}|\psi_i, n_i\rangle
\eeq
where the sum includes that over a basis $\psi_I$ of intermediate parent universe states.
The discussion readily generalizes to multiple species of BUs, and can be summarized by introducing operators $a_i,\, a^\dagger_i$ for the different species, together with different operators $\calo_i$ summarizing their couplings to the parent.

The  form of the BU couplings \eqref{BUham} implies that, in the free BU approximation, there is a simple relation between BU states and couplings.  Specifically, consider the states 
\beq
|\alpha\rangle = \mathcal{N}~e^{-\frac{1}{2}(a^\dagger - \alpha )^2} |0\rangle\ ;
\eeq
these diagonalize $a+a^\dagger$, with eigenvalue $\alpha$, and if we normalize them as $\mathcal{N}^2=e^{\alpha^2/2}/\sqrt{2\pi}$ then they satisfy the normalization convention
\beq
\langle \alpha|\alpha'\rangle = \delta(\alpha - \alpha')\ .
\eeq
Such a state is then an eigenstate of the hamiltonian  \eqref{BUham}, which takes the form
\beq
H_\alpha= H(\phi_i) + \alpha \int d^3x\, \calo(x)\ .
\eeq
Thus, the evolution is that of a theory with a new coupling constant multiplying the operator $\calo(x)$.  A more general BU state can be written as a superposition of the $\alpha$ eigenstates, and so can be thought of as describing an ensemble of such couplings.  This, together with weighting factors arising from disconnected parent universes, was for example proposed in \cite{ColeCC} to solve the (then) cosmological constant problem, by arguing that the weighting factors overwhelmingly prefer $\Lambda=0$.

\section{Renyis, replicas, and wormhole connections}\label{sec:renyisrepwhc}

We next turn to a discussion of what the rules of the wormhole calculus, combined with basic rules of quantum mechanics, imply in the context of computing quantities such as entropies that characterize the distribution of information in the system, as well as correlators.

\subsection{Entropies}

Suppose that one begins with an initial state $|\Psi_i\rangle$ for the combined parent/BU system, which in time $T$ then evolves by the hamiltonian \eqref{BUham} to
\beq\label{Tstate}
|\Psi, T\rangle = e^{-i H T}|\Psi_i\rangle\ ,
\eeq
with corresponding density matrix
\beq\label{totdens}
\rho(T) = |\Psi, T\rangle \langle\Psi, T|\ .
\eeq
A first simple problem is to compute Renyi entropies of this density matrix.  These are given by the standard formula
\beq\label{Renyis}
S_n = \frac{1}{1-n}\log \Tr (\rho^n)\ , 
\eeq
with the trace (over both parent and BU Hilbert spaces) given by
\beq\label{trtr}
\Tr(\rho^n) = \Tr\left(|\Psi, T\rangle \langle\Psi, T| \cdots |\Psi, T\rangle \langle\Psi, T|\right) = 1\ ,
\eeq
if states are properly normalized.
\begin{figure}[h!]
\begin{center}
 \hspace{-0.3cm} \begin{tikzpicture}[scale=1]
\pgftext{\includegraphics[scale=0.64]{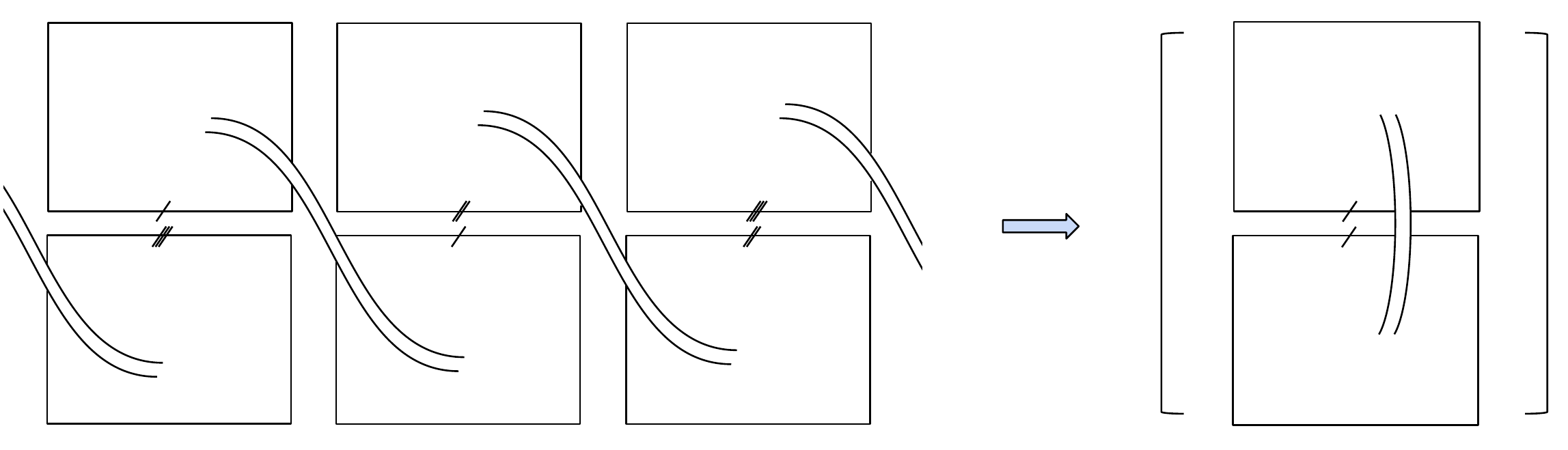}} at (0,0);
\draw (-6.83,-1.7) node  {\small $1$};
\draw (-6.83,1.7) node  {\small $\bar{1}$};
\draw (-4.05,-1.7) node  {\small $2$};
\draw (-4.05,1.7) node  {\small $\bar{2}$};
\draw (-1.25,-1.7) node  {\small $3$};
\draw (-1.25,1.7) node  {\small $\bar{3}$};
\draw (7.5,2) node  {\small $n$};
  \end{tikzpicture}
 \caption{\label{fig:whrep} Shown is a sketch of the geometry used in a replica method calculation of the 
 $n$th Renyi entropy of the  density matrix \eqref{totdens}. Time runs upwards (downwards) in the lower (upper) copies. This calculation produces only wormholes 
  that connect different replicas  in the pattern ${\bar 1}-2$, ${\bar 2}-3$, $\cdots$, ${\bar n} - 1$.  We also indicate how the parent universes  are identified at time $T$. The wormhole joining at $1$ is emitted from $\bar{n}$ while the one emitted from $\bar{3}$ joins $4$, etc. Wormholes connecting $1-1$, $\bar{1}-\bar{1}$, $2-2$, {\it etc.} are present, but not shown. The right panel shows a rearrangement of the diagrams making the purity of \eqref{totdens} manifest. }
\end{center}
\end{figure}
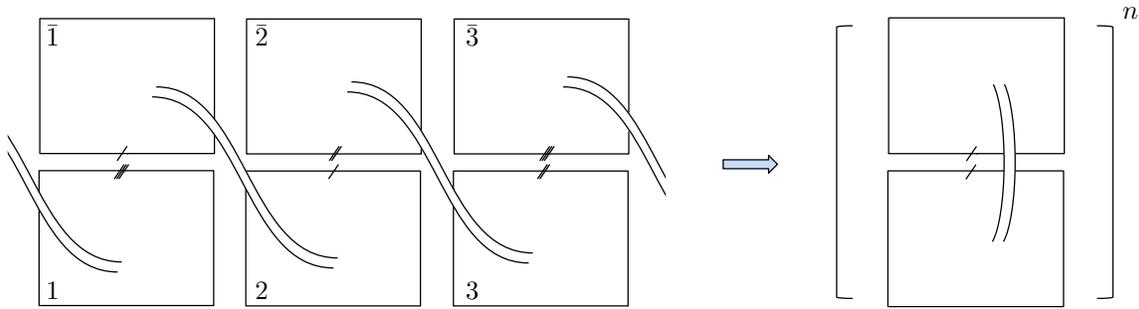

This seemingly trivial calculation already carries an important lesson regarding replicas and wormholes.  In the replica method\cite{CaWi}, each factor of $|\Psi, T\rangle \langle\Psi, T|$ may be represented in terms of a functional integral involving a replica copy of the geometry; the replica parent universes are pictured in fig.~\ref{fig:whrep}.  Then, when we calculate $\Tr (\rho^n)$, that implies a cyclic identification of final time slices of each factor, in the pattern ${\bar 1}-2$, ${\bar 2}-3$, $\cdots$, ${\bar n} - 1$. This {\it also} applies for the identification of the BUs of fig.~\ref{fig:whamp}. That is: {\it The rules of the wormhole calculus, combined with the standard quantum-mechanical rules for calculating the entropy $S_n$, imply wormhole connections only between neighboring replicas, in the preceding pattern -- they do {\rm not} imply that one should sum over geometries with wormhole connections between all replicas, in the way that is commonly conjectured\cite{Penington:2019kki, Almheiri:2019qdq}.}  

The wormhole calculus, together with standard quantum mechanical rules, dictate where replicas should be connected by wormholes.  Specifically, the wormhole connections follow from the contraction of indices between bra and ket factors, arising from either taking traces, or multiplying density matrices.
This principle is  expected to generalize to restrict replica topologies in cases where one has more complicated geometries contributing to amplitudes than simple BU emission/absorption. This conclusion does not change if we trace over a subregion of the parent universe (we comment on this below).

As one simple check, we show in fig.~\ref{fig:whrep} the pattern above allows us to rearrange the diagrams in a way that makes manifest that ${\rm Tr}(\rho^n)=({\rm Tr} \rho)^n$. This implies that $\rho$ is pure, which is consistent with the fact that we started from a pure state in the total Hilbert space (parent plus BUs) and the evolution is unitary. 

\begin{figure}[t!]
\begin{center}
\begin{tikzpicture}[scale=1]
\pgftext{\includegraphics[scale=0.4]{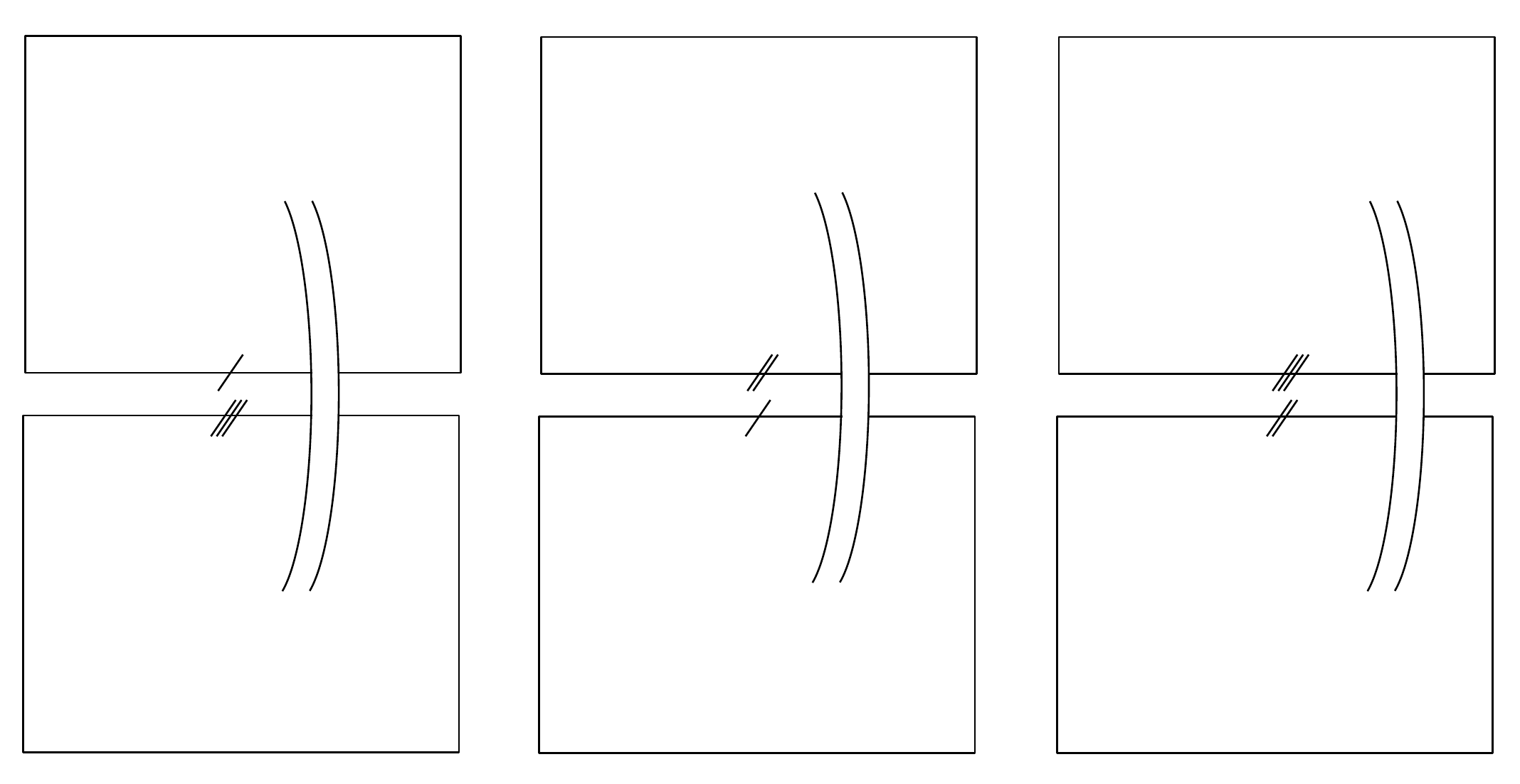}} at (0,0);
\draw (-3.9,-1.78) node  {\small $1$};
\draw (-3.9,1.75) node  {\small $\bar{1}$};
\draw (-0.95,-1.78) node  {\small $2$};
\draw (-0.95,1.75) node  {\small $\bar{2}$};
\draw (2,-1.78) node  {\small $3$};
\draw (2,1.75) node  {\small $\bar{3}$};
  \end{tikzpicture}
 \caption{\label{fig:whrep2}Shown is a sketch of the replica method calculation of the $n$th Renyi entropy of the reduced density matrix $\rho_p = {\rm Tr}_{\rm BU} \,\rho$. Time runs upwards (downwards) in the lower (upper) copies. Here there are no wormholes connecting different replicas, and the connections have the pattern ${\bar 1}-1$, ${\bar 2}-2$, $\cdots$, ${\bar n} - n$.  We also indicate how the parent universes are identified. Wormholes connecting $1-1$, $\bar{1}-\bar{1}$, $2-2$, {\it etc.} are present, but not shown.}
\end{center}
\end{figure}
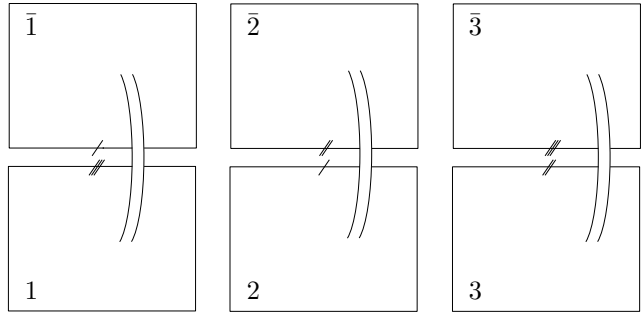

The preceding principle can be illustrated by a different calculation.  Suppose that we instead consider the density matrix of the parent universe,
\beq\label{pdens}
\rho_p = \Tr_{\rm BU} |\Psi, T\rangle \langle\Psi, T|\ ,
\eeq
and consider its Renyi entropies, given in terms of $\Tr(\rho_p^n)$.  The BU trace in \eqref{pdens} connects BUs in the bra and ket.  Then, when one calculates $\Tr(\rho_p^n)$, the final time slices on the parent universes are identified in the preceding pattern.  The BU connections instead form the pattern ${\bar 1}-1$, ${\bar 2}-2$, $\cdots$, ${\bar n} - n$, as illustrated in fig.~\ref{fig:whrep2}, 
but once again one does not sum over topologies with BUs connecting all replicas.  One also sees that, from the perspective of usual quantum mechanical rules, the latter kind of sum would appear rather unusual -- that would correspond to contracting various BU indices in a product such as \eqref{trtr} between {\it all} different factors. Fig.~\ref{fig:whrep2} also shows that due to the way the parent universe degrees of freedom are identified, now ${\rm Tr}(\rho_p^n)\neq({\rm Tr} \rho_p)^n$. This implies that $\rho_p$ is not pure, consistent with the QM interpretation. 

It is also informative to examine the corresponding expressions written in terms of the $\alpha$ vacua.  Consider, for example, an initial uncorrelated (product) state of the BUs and parent; after evolution by $T$, \eqref{Tstate} then gives
\beq\label{alphaevol}
|\Psi, T\rangle = \int d\alpha\, \psi(\alpha) U_\alpha(T) |\psi_i\rangle |\alpha\rangle\ ,
\eeq
where $U_\alpha(T) = \exp\{-i H_\alpha T\}$ is the evolution operator for a given $\alpha$.  Then $\rho_p$ becomes
\beq
\rho_p = \int d\alpha |\psi(\alpha)|^2\,  U_\alpha(T) |\psi_i\rangle \langle \psi_i| U_\alpha^\dagger(T)
\eeq
and the $n$th Renyi entropy is given by
\beq
\Tr(\rho_p^n) = \int \prod_{k=1}^n d\alpha_k |\psi(\alpha_k)|^2\, \langle \psi_i| U_{\alpha_k}^\dagger(T) U_{\alpha_{k+1}}(T) |\psi_i\rangle\ ,
\eeq
where we identify $\alpha_{n+1}=\alpha_1$.

In contrast, a sum over all possible wormhole connections between replicas (as suggested by \cite{Almheiri:2019qdq, Penington:2019kki}) would correspond to the expression
\beq\label{allconn}
\Tr(\rho_p^n) = \int d\alpha |\psi(\alpha)|^2\,  \left(\langle \psi_i| U_{\alpha}^\dagger(T) U_{\alpha}(T) |\psi_i\rangle\right)^n\ ,
\eeq
or, for evolution of an initial parent density matrix $\rho_{p,i}$, 
\beq\label{allconnrho}
\Tr(\rho_p^n) =  \int d\alpha |\psi(\alpha)|^2 \Tr\left(  U_{\alpha}(T)\rho_{p,i }U_{\alpha}^\dagger(T) \right)^n\ .
\eeq
While this behaves like an average over an ensemble of couplings with probability distribution $|\psi(\alpha)|^2$, it does not follow in a straightforward way from combining the rules for summing over topologies in amplitudes with a standard quantum-mechanical calculation.

We can use a similar analysis to treat the  entropy calculation for density matrices after tracing over excitations in a region $\mathcal{R}$ inside the parent universe. Specifically, we can compute the Renyi entropy associated to two different types of density matrix,
\begin{equation}
\rho^\mathcal{R} = {\rm Tr}_\mathcal{R}\, \rho\ ,~~~~~\text{and}~~~~~\rho_p^\mathcal{R} = {\rm Tr}_\mathcal{R}\, \rho_p\ .
\end{equation}
The first option does not include a trace over the BU Hilbert space while the second option does. 
When describing the wormhole connections coming from the sum over intermediate states, the $n$th Renyi entropy for the first and second options correspond to the connections shown in fig.~\ref{fig:whrep} and \ref{fig:whrep2} respectively. The only difference in the calculation is that now we will also identify the parent universe degrees of freedom corresponding to region $\mathcal{R}$ between the $i$th and $\bar{i}$th copies, with $i=1,\ldots, n$. The third option in which we sum over all possible wormholes (which is different from the quantum-mechanical rules we have been describing) is again  given by a sum over a single $\alpha$ parameter weighted by $|\psi(\alpha)|^2$, analogous to \eqref{allconnrho}.

One further comment is that the results in this section can be also reproduced using the methods of \cite{Klebanov:1988eh}. This is based on the fact that one can replace microscopic wormholes by a bilocal coupling between local operators inserted at their mouths. This interaction can be made local by introducing $\alpha$ parameters, which act as random coupling of local operators. In the cases studied above, when a wormhole is present between replicas, one can check (keeping track of combinatorics and phases) that the effect is to identify their $\alpha$ parameters. In the extreme case described last, where one includes all possible wormholes between any copies, the end result is to identify all $\alpha$ parameters, reproducing equation \eqref{allconnrho}.

\subsection{Correlators} 

\begin{figure}[t!]
\begin{center}
\begin{tikzpicture}[scale=1]
\pgftext{\includegraphics[scale=0.4]{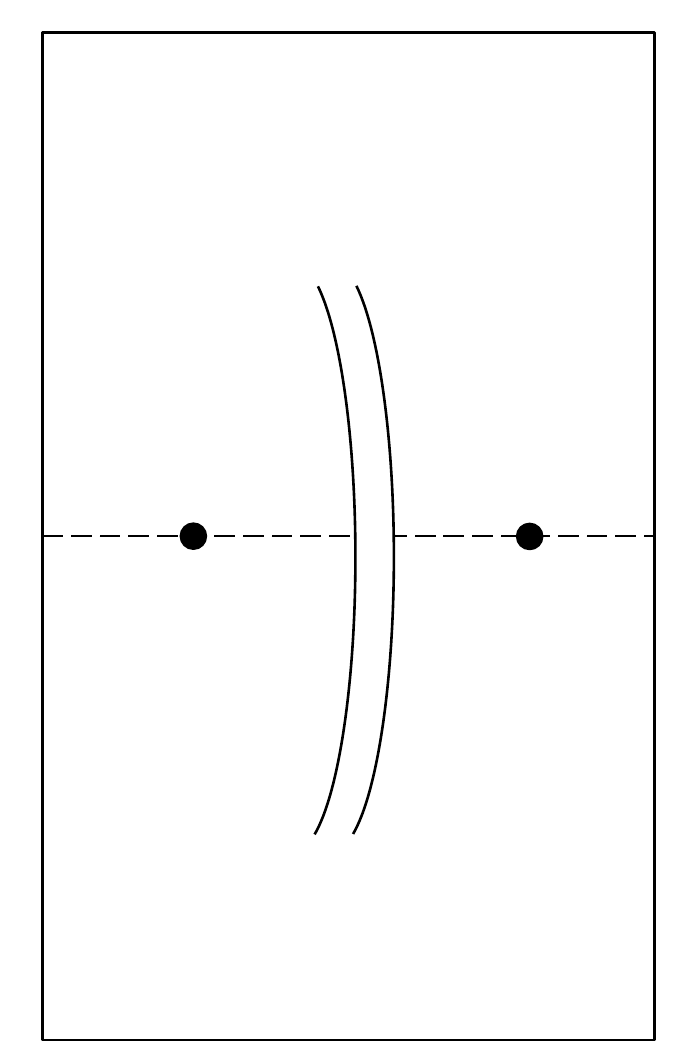}} at (0,0);
\draw (-1,-1.78) node  {\small $\Psi$};
\draw (-1,1.75) node  {\small $\bar{\Psi}$};
  \end{tikzpicture}
 \caption{\label{fig:whcorr} In this diagram we show the in-in calculation of a two point function (represented by the black dots). The bottom (top) represent the time evolution upwards (downwards) creating the bra (ket) at time $T$ along the dashed line, in a diagram like those described in \cite{GiSl3}. The state is glued at the dashed line including the operator insertions and the rules of QM would require us to include wormholes between them. We depict one of these wormholes.}
\end{center}
\end{figure}

As another application of these ideas we can analyze the computation of real time in-in correlators when summing over wormholes. We want to compute the expectation value of some operator $\mathcal{O}(\phi_i)$ acting on the parent universe degrees of freedom (but not acting on BU Hilbert space) at some time $T$. Then the rules of QM applied to this problem would give 
\begin{eqnarray}
\langle \mathcal{O}(\phi_i)\rangle &=& \langle \Psi, T | \mathcal{O}(\phi_i) | \Psi, T\rangle, \nonumber\\
&=& \int d\alpha |\psi(\alpha)|^2 \langle \psi_i | U_\alpha^\dagger(T) \mathcal{O}(\phi_i) U_\alpha(T) | \psi_i \rangle
\end{eqnarray} 
where we wrote the initial state as a linear combination of $\alpha$-states as in the previous section. The sum over intermediate states at time $T$, including BUs, imposes that the alpha parameters are  the same for the bra and the ket, giving the final formula above. Geometrically this comes from including wormholes between the bra and the ket as shown in fig.~\ref{fig:whcorr} (it is important in deriving this result that the operator does not act on the BU Hilbert space). Wormholes of a similar type that arise in calculating the density matrix were considered in \cite{Page:1986vw} (and more recently also in \cite{Maldacena:2019cbz}). 

In this example we see that thanks to the wormhole connecting the bra and ket the norm of the state is preserved under time evolution (if we set $\mathcal{O}(\phi_i) = 1$ then the evolution operators cancel). This would fail had we not included these wormholes, giving instead 
\beq
 \int d\alpha d\alpha' \hspace{0.08cm}\psi^*(\alpha)\psi(\alpha') \hspace{0.08cm} \langle \psi_i | U_\alpha^\dagger(T) \mathcal{O}(\phi_i) U_{\alpha'}(T) | \psi_i \rangle
\eeq
which does not preserve the norm under time evolution.

Some models of natural inflation are based on a non-perturbative axion potential generated by Euclidean wormholes \cite{Rey:1989mg, PhysRevLett.65.3233, Hebecker:2016dsw}. The considerations above would be relevant to compute, for example, the power spectrum or non-gaussianities in these models. 

\section{Determination of wormhole-induced couplings}

In \cite{Cole,GiStInc} it was argued that the growth of entropy that we see from the perspective of a parent universe if we begin in a generic BU state is not a good model for the kind of information loss originally proposed by Hawking to arise from black holes\cite{Hawking:1976ra}.   Specifically, models in \cite{Cole, GiStInc} showed that this information loss is not {\it repeatable}: if repeated experiments are performed, the entropy increase per experiment declines as their number increases.  This, together with the superselection rule for the $\alpha$ vacua, tell us that the entropy growth is associated with 
lack of knowledge of the specific value of the eigenvalue $\alpha$, or effective coupling constant, within the effective ensemble with probability distribution $|\psi(\alpha)|^2$.   We can use the preceding discussion to extend this argument, generalizing the argument  of \cite{Cole,GiStInc}, and also making contact with the question of replica wormholes.

Specifically, consider the evolution \eqref{alphaevol}, in the case where the parent universe wavefunction describes a number $s$ of independent systems, so
\beq
|\psi_i\rangle = |\tilde \psi_i\rangle^{\otimes s}\ .
\eeq
Suppose that these evolve as independent noninteracting systems (aside from wormhole connections), in which case  \eqref{alphaevol} takes the form
\beq
|\Psi, T\rangle = \int d\alpha\, \psi(\alpha) \left(\tilde U_\alpha(T) |\tilde \psi_i\rangle\right)^{\otimes s} |\alpha\rangle\ 
\eeq
with independent evolution operators $\tilde U_\alpha(T)$, and the parent density matrix becomes
\beq
\rho_{p,s}= \int d\alpha |\psi(\alpha)|^2\,  \left(\tilde U_\alpha(T) |\tilde \psi_i\rangle\right)^{\otimes s} \left(\langle \tilde \psi_i| \tilde U^\dagger_\alpha(T) \right)^{\otimes s}\ .
\eeq
The Renyi entropies are now given by
\beq\label{sReyn}
\Tr \rho_{p,s}^n = \int\prod_{k=1}^n  d\alpha_k | \psi(\alpha_k)|^2 \left( \langle \tilde \psi_i| \tilde U^\dagger_{\alpha_k}(T) U_{\alpha_{k+1}}(T) |\tilde \psi_i\rangle   \right)^s\ .
\eeq
At this stage, we again find that the underlying wormhole connections have a pattern like in \eqref{pdens} and in fig.~\ref{fig:whrep2}, once again connecting $1-{\bar 1},\cdots,n-{\bar n}$.

To evaluate the Renyi entropies for large $s$, note that
the inner products in \eqref{sReyn} can be written
\beq
\langle \tilde \psi_i| \tilde U^\dagger_{\alpha}(T) \tilde U_{\alpha'}(T) |\tilde \psi_i\rangle = e^{\gamma(\alpha,\alpha')+i \delta(\alpha,\alpha')}\ ,
\eeq
with real functions $\gamma$ and $\delta$; we have $\gamma(\alpha,\alpha)=\delta(\alpha,\alpha)=0$, $\delta(\alpha',\alpha)= -\delta(\alpha,\alpha')$, and generically $\gamma(\alpha,\alpha')<0$ for $\alpha\neq\alpha'$.  This means that the integrals in \eqref{sReyn} become increasingly sharply peaked at $\alpha_k=\alpha_{k+1}$ for large $s$.  
Near $\alpha=\alpha'$, we have  expansions $\gamma(\alpha,\alpha') = -C(\alpha-\alpha')^2 +\cdots$ and $\delta(\alpha,\alpha') = D(\alpha-\alpha') + E(\alpha-\alpha')^3 + \cdots$.  Inserting these in \eqref{sReyn}, the $D$ terms cancel, the $E$ terms contribute at subleading order in $1/s$, and we find
\beq
\Tr \rho_{p,s}^n \approx \int\prod_{k=1}^n  d\alpha_k | \psi(\alpha_k)|^2  e^{-sC(\alpha_k-\alpha_{k+1})^2} \ .
\eeq
For a large number  $s$ of experiments, the form of the integrals is determined by the $n-1$ gaussian factors with width $\sim 1/\sqrt s$ (excluding an overall ``center of mass" integral), and so the entropies become
\beq
\Tr\, \rho_{p,s}^n \approx \left(\frac{1}{\sqrt s}\right)^{n-1} F(n), 
\eeq
for some function $F(n)$ with $F(1)=1$.  
The Renyi entropies \eqref{Renyis} then are
\beq
S_n(s) \approx \frac{1}{2} \log s + \frac{1}{1-n} \log F(n)\ ,
\eeq
and the change of a given Renyi entropy per experiment is 
\beq
\frac{d}{ds} S_n(s) \approx \frac{1}{2s}\ .
\eeq

In summary, there is a ``loss of information" in subsequent experiments conducted in the parent universe, which is associated with the lack of information about the state of the BUs.  However, since the BU state effectively mimics a probability distribution for coupling constants, successive experiments better and better determine the a-priori uncertain values of these couplings.  In the limit of a large number of experiments, $s\rightarrow\infty$, the indeterminacy vanishes, and there is no further growth of entropy/loss of information.

\section{Discussion and lessons}

As was first shown in \cite{Cole,GiStInc}, the effect of BUs is to contribute to coupling constants multiplying  operators that summarize the effect of a given kind of BU on the fields of a parent universe.  A generic state of BUs leads to a probability distribution for such couplings. We have shown here, generalizing arguments in \cite{Cole,GiStInc}, that successive experiments lead to an increasingly precise determination of such couplings, such that in the limit of a large number of experiments, additional experiments experience no further loss of information.  One can think of this determination process as a ``collapse of the wavefunction" into an $\alpha$ state of the BUs corresponding to a particular set of couplings.  There is a well-developed set of rules, the wormhole calculus,\cite{GiStInc},\cite{Cole} that were overviewed in section two and underpin this set of observations.  There are effects that go beyond the simple free BU approximation used there, and account for interactions between BUs and with other parent universes; an initial account of such effects in a more general third-quantized approach was given in \cite{GiSt3Q}, and some such effects were argued to lead to specific distributions effectively fixing couplings such as the cosmological constant in \cite{ColeCC}.

One may calculate quantities such as entropies and correlators, in the presence of topology change/BUs, and within the framework of the wormhole calculus, using standard quantum-mechanical rules for doing so.  In particular, the wormhole calculus may be combined with replica methods\cite{CaWi}.
When one does this, the standard quantum mechanical rules applied to the entropies or correlators we consider
lead to a limited pattern of wormhole connections between replicas.  These for example only produce a connection between replicas that are ``nearest neighbors," and do not produce connections between different ``bra" copies or ``ket" copies.

The work of  \cite{Penington:2019kki, Almheiri:2019qdq} considers even more general topologies that go outside of these nearest neighbor and bra-ket constraints.  An important question is how to justify such connections, based on an underlying consistent set of rules for computing amplitudes including topology change, and following standard quantum rules, {\it e.g.} based on tracing over appropriate states, for sewing amplitudes.  In fact, given that the replica wormhole configurations of \cite{Penington:2019kki, Almheiri:2019qdq} involve even more complicated topological connections between replicas than combinations of single-wormhole connections, their interpretation in terms of traces of appropriate density matrices seems even more obscure.
There seem to be at least three different possibilities for explaining a role for such extended rules for replica connections.  One is that they correspond to calculating other more general mathematical quantities than the simple entropies one usually considers.  A second is that they represent some modification of the usual quantum-mechanical rules for composing amplitudes.  A third is that they give an effective parameterization of other effects that are directly described without invoking such extended rules.  It does appear, as seen in \eqref{allconn}, \eqref{allconnrho}, that some such expressions can describe certain ensemble averages for BU couplings.

If topological or BU effects do help in resolving the unitarity crisis associated with BH evolution, a key question is to understand the underlying transition amplitudes describing how they do so.  In particular, as discussed in section four, one may perform a large number of experiments, after which evolution in our parent universe should be unitary, with no further loss of information.  Once we have ``collapsed the BU wavefunction" in this fashion, we can then consider subsequent scattering experiments where BHs form and decay, and those processes should be described by unitary amplitudes.  However, at this stage the net effect of the BUs is, at least neglecting higher-order effects ({\it e.g.} as in \cite{GiSt3Q}), simply to contribute to various coupling constants.  In the resulting effective theory, we can then ask how BH formation and decay evades the standard information-loss arguments going back to Hawking's original work \cite{Hawking:1974sw}.

It has been argued that what is needed for such a unitary description are interactions that, when viewed from the perspective of an effective field theory description of BH evolution, transfer information or entanglement from the internal state of the BH to the environment of the BH\cite{NLvC,SGmodels,BHQIUE,Giddings:2012dh, Giddings:2013kcj, Giddings:2012gc}.  In particular, Refs.~\cite{NVU,BHQU} give a parameterization of such interactions in an effective theory.  One possibility is that the topology-changing processes somehow contribute to such interactions, which appear to be nonlocal from the effective field theory perspective.  We have seen that simple connections, via a small spacetime wormhole, between two different points do not induce the right kind of nonlocal transfer of information, but possibly contributions of larger wormholes, say comparable to the BH size, could, as has been suggested in \cite{Saad:2019lba,Penington:2019kki} and \cite{Marolf:2020xie}.  If this were the case, such interactions could likely be parameterized in the general framework of \cite{NVU,BHQU}.  However, in order to justify such a picture, and even more importantly, to give a precise description of such interactions that allows one to calculate the effects on outgoing fields (and on possibly observable quantities), one needs a description of how the topology-changing processes contribute to amplitudes.  This might, for example, involve finding instanton-like or other similar processes operating on scales comparable to a BH's size.  A preliminary investigation reveals a number of subtleties in giving any systematic description of such effects\cite{GiTuinprog}, but it is worth determining whether progress can be made in this direction.

\vskip.3in
\noindent{\bf Acknowledgements} We wish to thank J. Maldacena, D. Marolf, and H. Maxfield for useful discussions. This material is based upon work supported in part by the U.S. Department of Energy, Office of Science, under Award Number {DE-SC}0011702; the work of GJT is also supported by a Fundamental Physics Fellowship.

\appendix

\section{Puzzles for replica wormholes}

In \cite{Penington:2019kki, Almheiri:2019qdq}, it has specifically been argued that replica wormholes make important contributions to the entropy, such that it follows the Page curve corresponding to unitary evolution.  It is important to better understand these arguments, and in particular the question of how they might relate  to underlying unitary quantum amplitudes describing black hole formation and evaporation.  Since the main body of the paper has shown how amplitudes including wormholes may be combined into density matrices, and then entropies may be computed from density matrices, following basic quantum-mechanical rules for composing amplitudes and taking traces, a specific question is whether one can ``reverse engineer" the geometries of 
\cite{Penington:2019kki, Almheiri:2019qdq} to infer the structure of underlying quantum amplitudes.  In particular, this can be asked in the context of the specific rules we have found for replica geometries.  One new ingredient of this discussion is that it goes beyond the free approximation.  We find some interesting puzzles regarding possible connection to underlying amplitudes.

\begin{figure}[h!]
\begin{center}
 \begin{tikzpicture}[scale=1]
\pgftext{\includegraphics[scale=0.6]{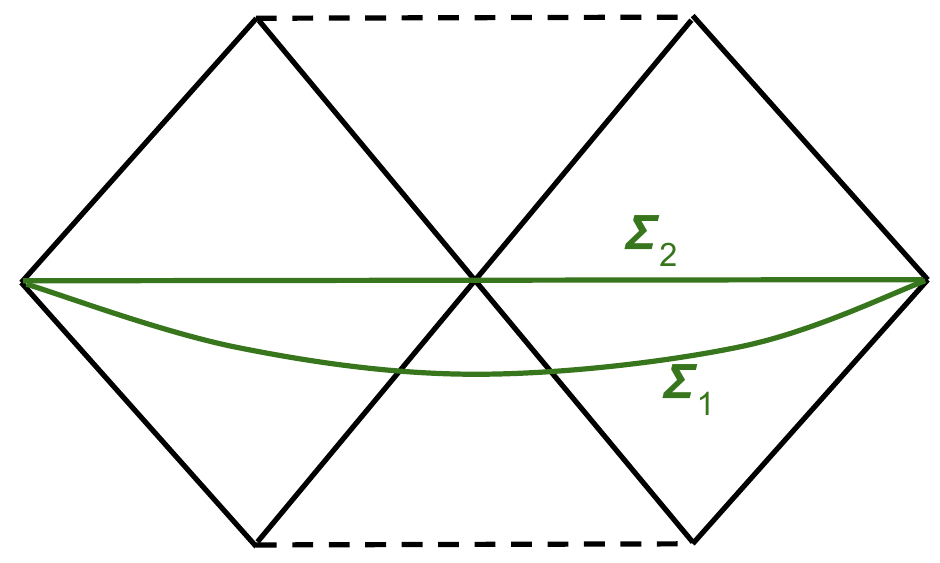}} at (0,0);
\draw (0,-2.2) node  {\small $(a)$};
  \end{tikzpicture}
  \hspace{1cm} \begin{tikzpicture}[scale=1]
\pgftext{\includegraphics[scale=0.7]{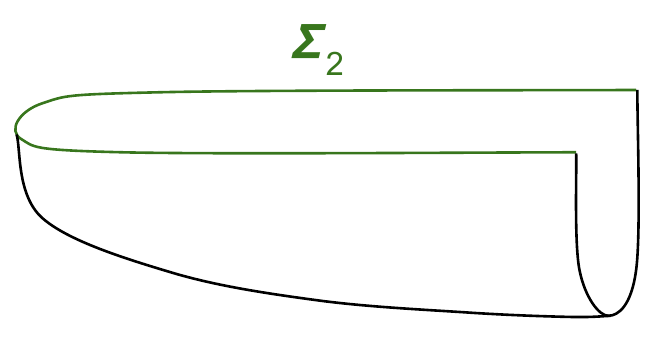}} at (0,0);
\draw (0,-2.2) node  {\small $(b)$};
  \end{tikzpicture}
 \caption{\label{fig:appendix1} (a) Penrose diagram of a Schwarschild black hole. We indicate two Cauchy slices in the two-sided spacetime, $\Sigma_1$ and $\Sigma_2$. (b) Euclidean no-boundary preparation of the two-sided state on the surface $\Sigma_2$. }
\end{center}
\end{figure}

As a warmup, first consider amplitudes corresponding to states of the two-sided black hole, as shown in Fig.~\ref{fig:appendix1}; for simplicity we focus on a Schwarzschild black hole. We can for example  describe a state on the Cauchy surface $\Sigma_2$ by evolving from another surface $\Sigma_1$ at a previous time. This state is a functional $\Psi_{\Sigma_2}(g, \phi,\ldots)$, of the metric $g$ on $\Sigma_2$,  with $\phi$ denoting possible matter fields.  A particularly simple state is prepared on $\Sigma_2$ through the Hartle-Hawking prescription, as pictured in Fig.~\ref{fig:appendix1}(b).  If the latter is viewed as a contribution to the sum over geometries, it can be thought of describing production of a two-sided black hole from ``nothing."  A more clearly motivated version of this geometry is when the two-sided black hole is magnetically charged; instantons that describe the Schwinger pair production of such black holes in a background magnetic field have been described in \cite{Gibbons:1986cq,Garfinkle:1990eq,Garfinkle:1993xk}, and have near-horizon structure of precisely the same form (see, {\it e.g.}, Fig.~1 of \cite{Garfinkle:1993xk}).

\begin{figure}[h!]
\begin{center}
 \begin{tikzpicture}[scale=0.7]
\pgftext{\includegraphics[scale=0.3]{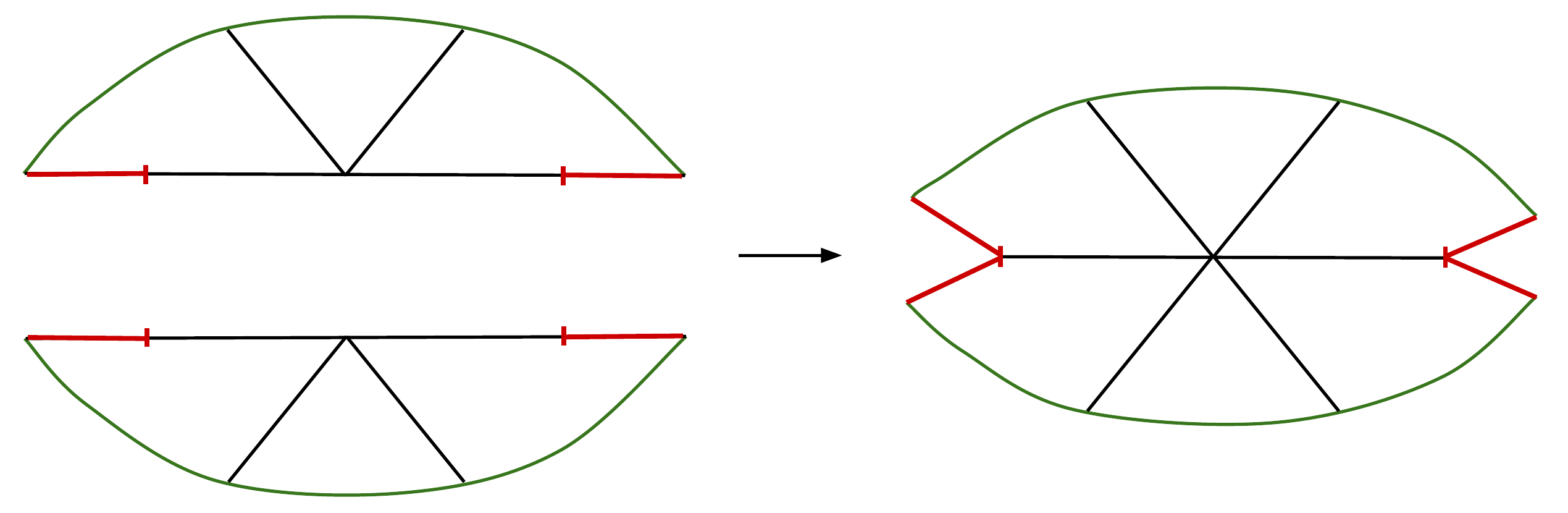}} at (0,0); 
\draw (-4.8,0) node  {\small $\rho_{\hspace{0.02cm}\mathcal{R}}~\sim$};
\draw (-3.5,0) node  {\small $\mathcal{R}$};
\draw (-0.8,0) node  {\small $\mathcal{R}$};
\draw (-2.2,0) node  {\small sew};
\draw (0,-2) node  {\small $(a)$};
  \end{tikzpicture}
 ~~~ ~ \begin{tikzpicture}[scale=0.7]
\pgftext{\includegraphics[scale=0.3]{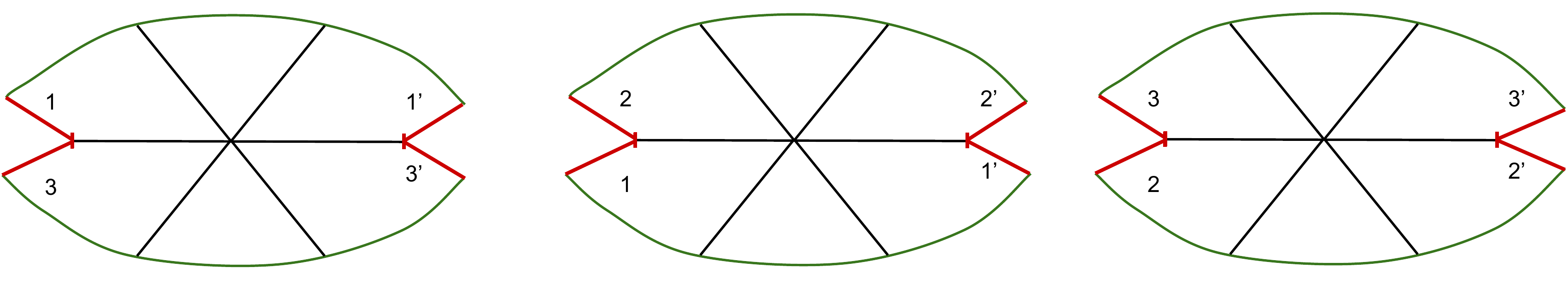}} at (0,0); 
\draw (-6.55,0.05) node  {\small ${\rm Tr}\hspace{0.05cm} \rho_{\mathcal{R}}^3~\sim$};
\draw (0,-2) node  {\small $(b)$};
  \end{tikzpicture}
 \caption{\label{fig:appendixRho1} (a) Using the Lorenzian preparation of the state (region between $\Sigma_1$ - green curve - and $\Sigma_2$ - black and red line) on slice $\Sigma_2$ this picture denotes the operation that computes the partial trace of the density matrix $\rho_{\mathcal{R}}$ (where $\mathcal{R}\subset \Sigma_2$ are the segments in red) after tracing over the complementary region $\mathcal{R}_c$. (b) Computation of the third Renyi entropy. The states on the red segments corresponding to the copies of $\mathcal{R}$ are identified as indicated by the numbers.}
\end{center}
\end{figure}
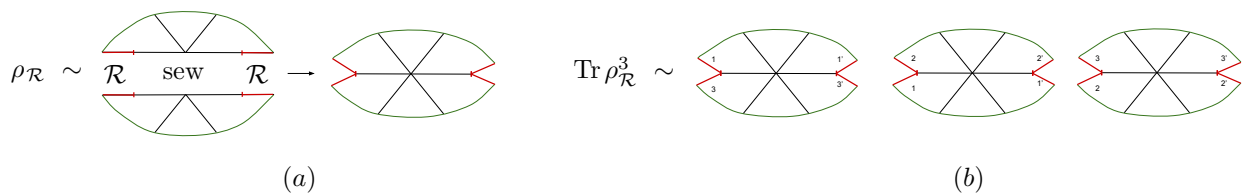

Having prepared a state by one of these methods, we can calculate the entropy of a subregion $\mathcal{R}$ outside the black hole, with corresponding  density matrix $\rho_{\mathcal{R}}= {\rm Tr}_{\mathcal{R}_c} |\Psi \rangle \langle \Psi |$, where $ \mathcal{R}_c$ is the complement of the region $\mathcal{R}$ and contains the black hole. The growth of entropy with time in this region follows the original ``Hawking" curve even past the Page time, when the calculation is done in the spacetimes shown in Fig.~\ref{fig:appendix1}. This density matrix can be represented by the diagram Fig.~\ref{fig:appendixRho1}(a), where we used the Lorenzian preparation of the state shown in Fig.~\ref{fig:appendix1}(a). The Renyi entropy calculation is described by Fig.~\ref{fig:appendixRho1}(b) with the identifications shown, sewing the boundaries along the identified copies of $\mathcal{R}$. 

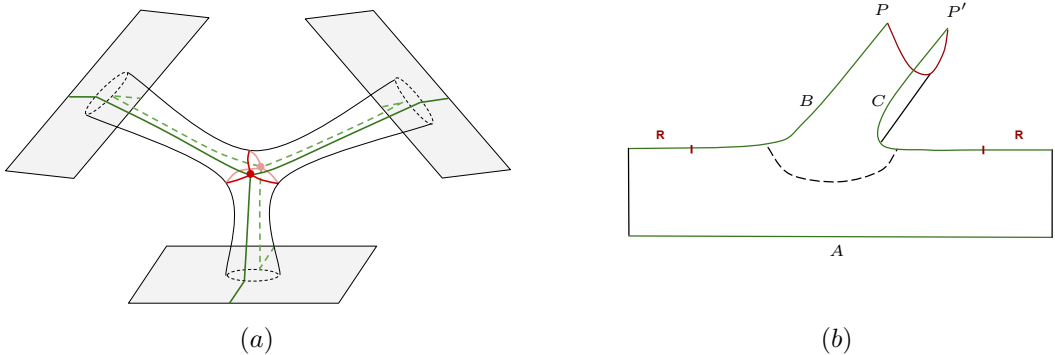
\begin{figure}[h!]
\begin{center}
 \begin{tikzpicture}[scale=1]
\pgftext{\includegraphics[scale=0.4]{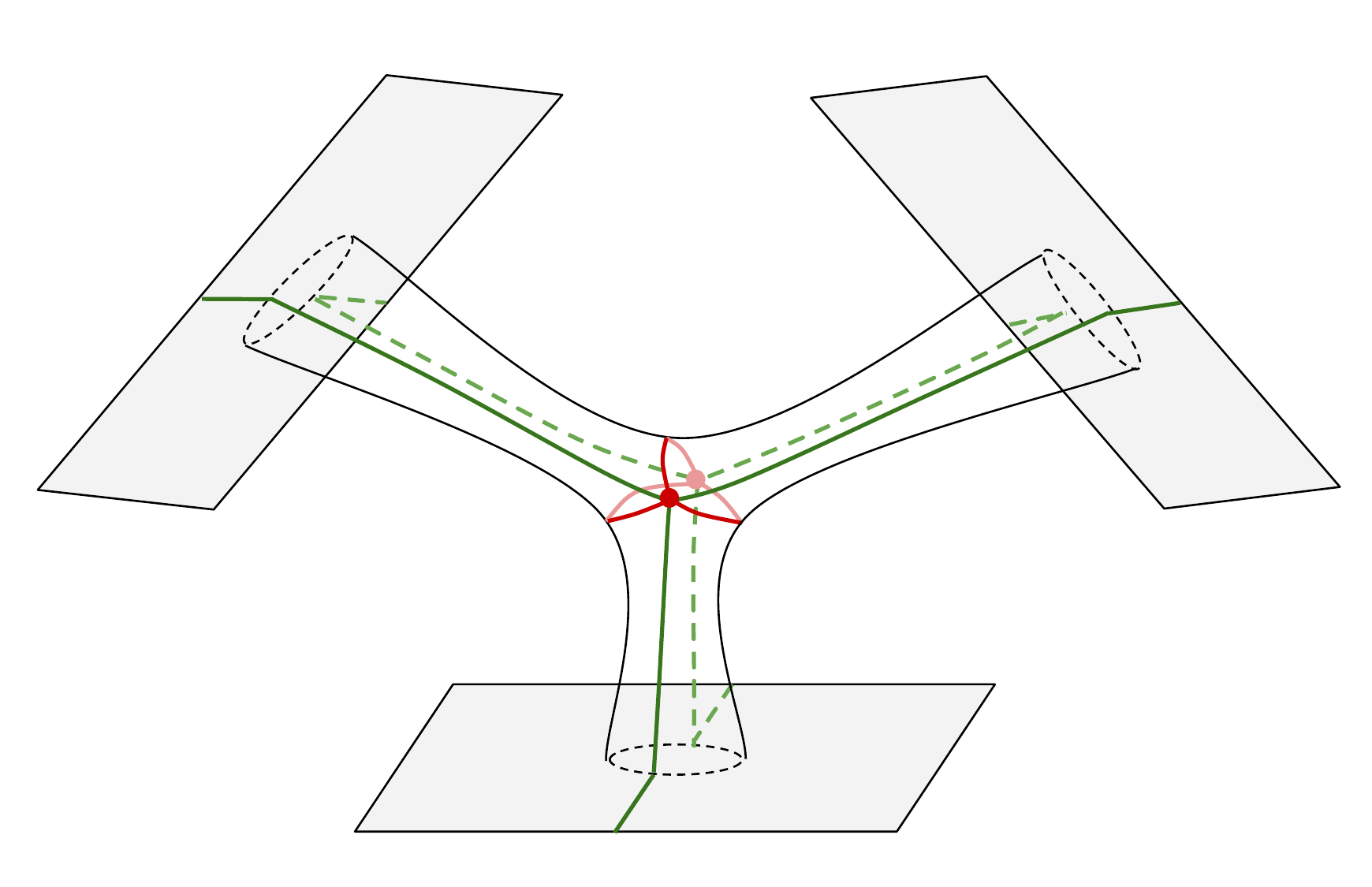}} at (0,0);
\draw (0,-2.5) node  {\small $(a)$};
  \end{tikzpicture}
   \hspace{1cm}\begin{tikzpicture}[scale=0.8]
\pgftext{\includegraphics[scale=0.4]{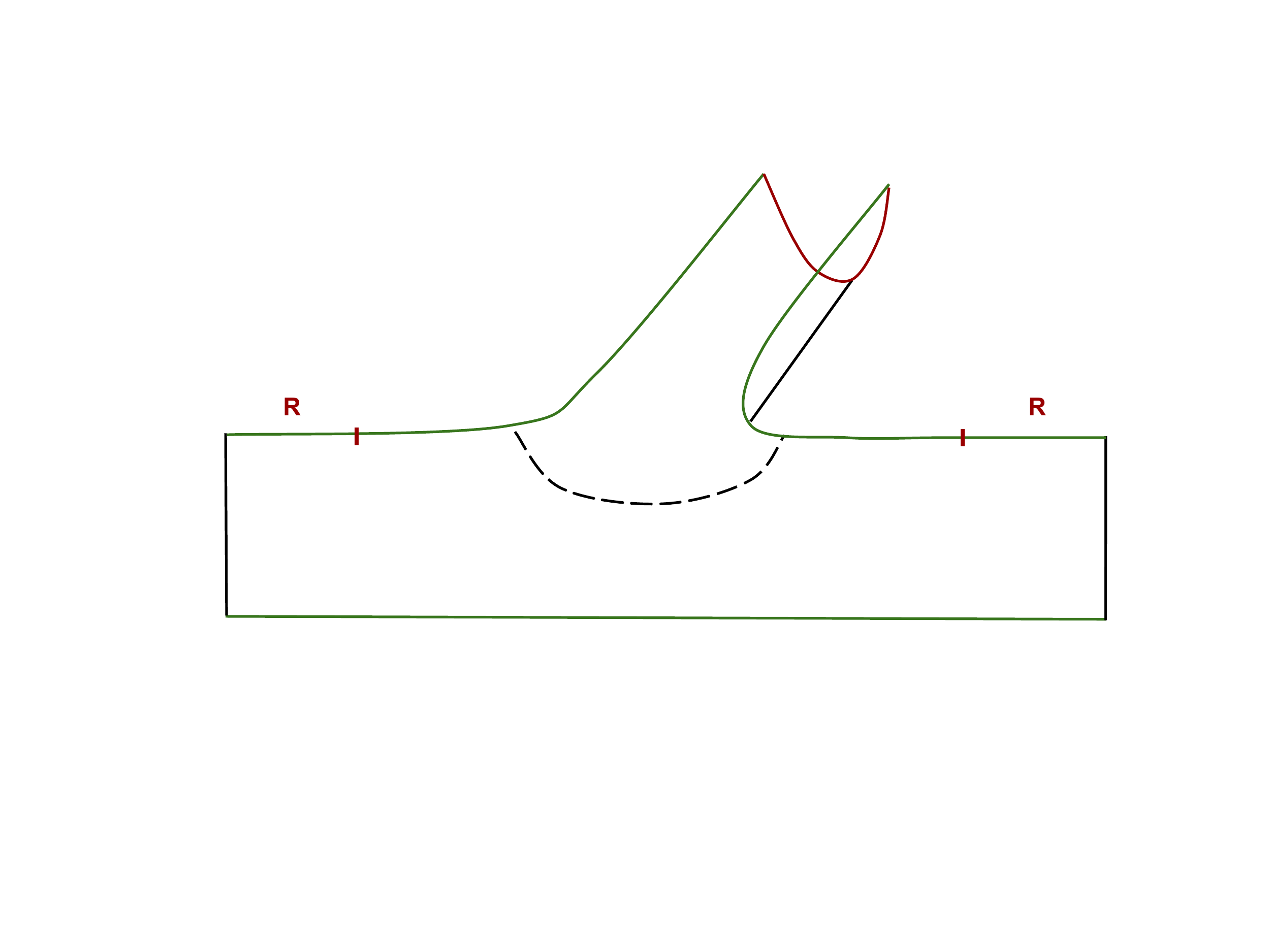}} at (0,0);
\draw (0.75,2) node  {\tiny $P$};
\draw (2,2) node  {\tiny $P'$};
\draw (-0.5,0.5) node  {\tiny $B$};
\draw (0.7,0.5) node  {\tiny $C$};
\draw (0,-2) node  {\tiny $A$};
\draw (0,-3.5) node  {\small $(b)$};
  \end{tikzpicture}
 \caption{\label{fig:pinwheel} (a) Pinwheel amplitude for three replicas. The green lines are taken to be constant time slices. The replicas are connected through a branch cut along $\mathcal{R}$ that we do not draw to avoid clutter. (b) Building block. Constant time slice is the parent universe (green slice); the possible interpretation of the red segment is discussed in the text. We also indicate the region $\mathcal{R}$ where radiation is collected.}
\end{center}
\end{figure}

Bearing these preliminary examples in mind, we would like to understand the interpretation of the replica wormholes of \cite{Penington:2019kki, Almheiri:2019qdq} which were argued to correct the Hawking curve and produce the Page curve.  An important contribution was from the ``pinwheel geometry," which for the third Renyi entropy is drawn in Fig.~\ref{fig:pinwheel}(a).  Since we have seen how black hole geometries may be sewn to calculate entropies, the question is how this diagram may be unstitched to describe underlying amplitudes.

There are two ways to try to interpret the pinwheel geometry, and both seem to be incompatible with a quantum mechanical treatment of a parent plus baby universe Hilbert space. One approach is to view this geometry as arising from a baby universe interaction. In this interpretation each replica creates its own baby universe and the central part of the picture represents their interaction vertex. But, interactions between all replicas simultaneously are not consistent with the rules developed in section \ref{sec:renyisrepwhc}, even outside the free approximation.  

A second possibility arises from cutting the pinwheel along the green lines in Fig.~\ref{fig:pinwheel}(a).  This is analogous to unstitching the previous figures, Fig.~\ref{fig:appendixRho1}(a) and (b), along the corresponding black lines, to return to the underlying amplitudes.  If we do this, for each replica, 
then the pinwheel can be understood as gluing six copies of the building block shown in Fig.~\ref{fig:pinwheel}(b).

The upper green boundary in Fig.~\ref{fig:pinwheel}(b) is identified with the corresponding bra amplitude, i.e. $\bar{1}-1$, $\bar{2}-2$, {\it etc.}, inside the region $\mathcal{R}_c$. The red boundary is identified instead with a different replica. Note that, if one is summing over all geometries, this identification can be to a {\it different} replica than the corresponding replica identifications on region $\mathcal{R}$, analogous to those of Fig.~\ref{fig:appendixRho1}(b).  This could be made more consistent with the rules given in section \ref{sec:renyisrepwhc} if the connections are given between consecutive replicas. 

Even with this assumption, puzzles remain. Specifically, 
the interpretation of the underlying geometry of Fig.~\ref{fig:pinwheel}(b), is unclear;  it is not obvious what lorentzian amplitude this contributes to.  First, ignoring the red segment, it appears that this describes a transition from a two-sided black hole, with spatial geometry labeled $A$ (bottom green line), to two separate spatial components $B$ and $C$ (top separate green segments). That leaves the question of interpreting the red segment, which is glued to a different replica.  If it is not interpreted as part of the geometry of the final slice, the latter geometry remains disconnected.  If, instead, it is interpreted as part of that geometry, it is not clear how to understand these gluing conditions.  Indeed, the points $P, P'$ at the junction between this segment and the components $B$ and $C$ are common to {\it all} the replicas, as seen in Fig.~\ref{fig:pinwheel}(a).  It is
 not clear how such a gluing prescription is compatible with the rules of
 section \ref{sec:renyisrepwhc}. 

We believe it is important to understand what  unitary amplitudes, if any, underly the formal entropy calculations of \cite{Penington:2019kki, Almheiri:2019qdq}, if the underlying calculational rules can be understood without modifying quantum mechanics.  The present work outlines a na\"\i ve attempt to interpret these, which leads to puzzles.  We leave other efforts to understand these to future work.

\mciteSetMidEndSepPunct{}{\ifmciteBstWouldAddEndPunct.\else\fi}{\relax}
\bibliographystyle{utphys}
\bibliography{WHC}{}

\end{document}